\documentclass[Physsubmission, Phys]{SciPost}

\graphicspath{{figures/}}
\usepackage{siunitx}

\def \rinv {$R_{\mathrm{inv}}$}

\def \CT {$C_{2}$}
\def \tauTO {$\tau_{21}$}
\def \tauTT {$\tau_{32}$}

\def \Mphi {M$_{\phi}$}

\hypersetup{
    colorlinks,
    linkcolor={red!50!black},
    citecolor={blue!50!black},
    urlcolor={blue!80!black}
}

\usepackage[bitstream-charter]{mathdesign}
\DeclareSymbolFont{usualmathcal}{OMS}{cmsy}{m}{n}
\DeclareSymbolFontAlphabet{\mathcal}{usualmathcal}

\begin{document}

\begin{center}{\Large \textbf{
Towards better discrimination and improved modelling of dark-sector showers\\
}}\end{center}

\begin{center}
Andy Buckley\textsuperscript{1},
Deepak Kar\textsuperscript{2} and
Sukanya Sinha\textsuperscript{3$\star$}
\end{center}

\begin{center}
{\bf 1} School of Physics and Astronomy, University of Glasgow, Glasgow, Scotland
\\
{\bf 2} School of Physics, University of Witwatersrand, Johannesburg, South Africa
\\
{\bf 3} School of Physics, University of Witwatersrand, Johannesburg, South Africa
\\
* sukanya.sinha@cern.ch
\end{center}

\begin{center}
\today
\end{center}


\definecolor{palegray}{gray}{0.95}
\begin{center}
\colorbox{palegray}{
  \begin{tabular}{rr}
  \begin{minipage}{0.1\textwidth}
    \includegraphics[width=23mm]{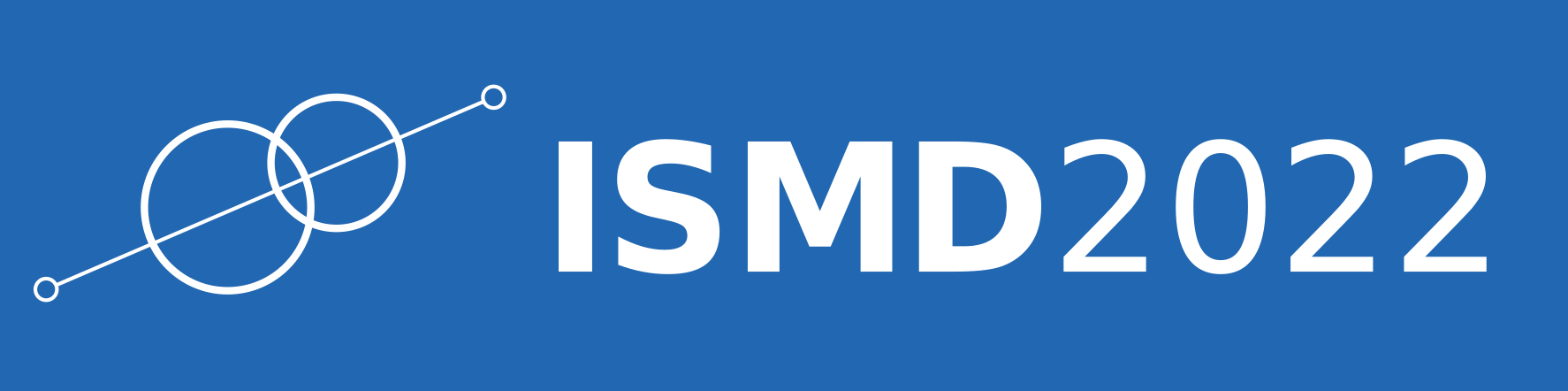}
  \end{minipage}
  &
  \begin{minipage}{0.8\textwidth}
    \begin{center}
    {\it 51st International Symposium on Multiparticle Dynamics (ISMD2022)}\\ 
    {\it Pitlochry, Scottish Highlands, 1-5 August 2022} \\
    \end{center}
  \end{minipage}
\end{tabular}
}
\end{center}

\section*{Abstract}
{\bf
As no evidence for classic WIMP-based signatures of dark matter have been found at the LHC, several 
phenomenological studies have raised the possibility of accessing a strongly-interacting dark sector through 
new collider-event topologies. If dark mesons exist, their evolution and hadronization procedure are currently 
little constrained. They could decay promptly and result in QCD-like jet structures, even though the original 
decaying particles are dark sector ones; they could behave as semi-visible jets; or they could behave as 
completely detector-stable hadrons, in which case the final state is just the missing transverse momentum. In 
this contribution we will introduce a study performed to explore use of jet substructure methods to distinguish 
dark-sector from QCD jets in the first two scenarios, using observables in a IRC-safe linear basis, and discuss 
ways forward for this approach to dark-matter at the LHC.
}

\vspace{10pt}
\noindent\rule{\textwidth}{1pt}
\tableofcontents\thispagestyle{fancy}
\noindent\rule{\textwidth}{1pt}
\vspace{10pt}

\section{Introduction}
\label{sec:intro}
In recent years, there has been an increase in the number of search programmes exploring the possiblity of a 'dark sector' beyond the Standard Model (BSM) using LHC data. To date, dark-matter searches at the LHC have usually focused on Weakly Interacting Massive Particles (WIMPs), but since the standard signatures have found no compelling evidence, several recent phenomenology papers~\cite{Cohen:2015toa,Cohen:2017pzm,Park:2017rfb} have explored the possibility of accessing the dark sector with unique collider topologies. If dark mesons exist, their evolution and hadronization procedure are currently little constrained. They could decay promptly and result in a very Standard Model (SM) QCD-like jet structure, even though the original decaying particles are dark-sector ones; they could behave as semi-visible jets; or they could behave as completely detector-stable hadrons, in which case the final state is just the missing transverse momentum. Apart from the last case, which is more like a conventional BSM missing transverse-momentum (MET) signature, the modelling of these scenarios is somewhat an unexplored area, other than the range of phenomenological predictions as implemented in Pythia8's HV module~\cite{Sjostrand:2014zea,Carloni:2010tw}. 

This contribution explores the possibility of designing observable(s) to distinguish between dark jets, semi-visible jets and light quark/gluon jets by comparing different observables. Examples are combinations of jet substructure variables like energy correlation functions~\cite{Jankowiak:2012na,Larkoski:2013eya,Moult:2016cvt} (ECFs) and $n$-subjettiness~\cite{Thaler:2010tr}, which characterise moments of the energy/particle distributions within jets. These families of variables feature angular scaling parameters that vary their sensitivity to different angular scales of jet emissions, potentially sensitive to the changes in jet structure introduced by dark-shower splitting, for various dark-hadron masses.

There have been some studies of looking at standalone JSS observables with focus on energy correlation observables~\cite{Cohen:2020afv,Kar:2020bws} that discussed the non trivial theoretical uncertainties associated with jet substructure. Studying the variations of such observables, and their uncertainties between MC models and MC theory systematic uncertainties, will enable a comprehensive survey of how to maximise measurement sensitivity across the BSM model space; in particular, varying the DM mass might affect the shower, since it impacts the semi-visible splitting kinematics, and allow to design “theory-safe” variables directly motivated by the splitting structure. 


\section{Event generation}
The signal samples, at $\sqrt{s}=13$ TeV are generated by using a t-channel simplified dark-matter model in Madgraph5~\cite{Alwall:2014hca} matrix element (ME) generator, with $xqcut = 100$ \footnote{defined as the minimum $k_{\textrm{T}}$ separation between partons}
and NNPDF2.3 LO PDF set~\cite{Skands:2014pea}, a mediator mass of 1500~GeV, and a dark-matter candidate mass of 10~GeV. Different \rinv\ fractions result in somewhat different kinematics, so \rinv\ values of 0.3, and 0.7 are studied.
The process $p p \rightarrow{\chi\bar{\chi}}$ with up to two extra jets were simulated and MLM matched~\cite{Mangano:2001xp} to have a reasonable cross-section and obtain a proper signal which does not get swamped under multijet background. The 
multijet production described by QCD are generated with Pythia8.

\section{Exploring new observables for dark-sectors}
Energy flow polynomials~\cite{Komiske:2017aww} (EFPs) are observables that are multi-particle energy correlators with specific angular structures which directly result from IRC safety. EFPs form a linear basis of all IRC-safe observables, making them suitable for a wide variety of jet substructure contexts where linear methods are applicable. 
For a multigraph $G$ with $N$ vertices and edges $(k, l) \subset G$, the corresponding EFP takes the form
\begin{equation}
    \mathrm{EFP_{\mathrm{G}}} = \Sigma^M_{i_1 = 1} ... \Sigma^M_{i_N = 1} z_{i_1} ... z_{i_N} \Pi_{(k, l) \subset G} \theta_{i_k i_l} \, ,
\end{equation}
where the jet consists of $M$ particles, $z_i$ is the energy fraction carried by particle $i$, and $\theta_{ij}$ is the angular distance between particles $i$ and $j$.

\begin{figure}[ht]
  \centering
  \includegraphics[width=0.8\textwidth]{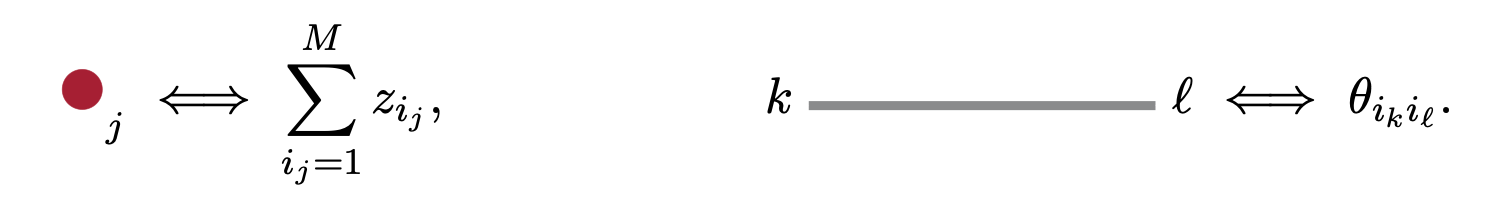}\\
  \caption{EFP construction: vertex and angular connectors}
  \label{fig:efpdiag}
\end{figure}

Each edge $(k,l)$ in a multigraph is in one-to-one
correspondence with a term $\theta$ in an angular
monomial. Each vertex $j$ in the multigraph corresponds to a
factor of $z$ and summation over $i_j$ in the EFP, as can be seen from Figure~\ref{fig:efpdiag}.

\begin{figure}[ht]
  \centering
  \includegraphics[width=0.5\textwidth]{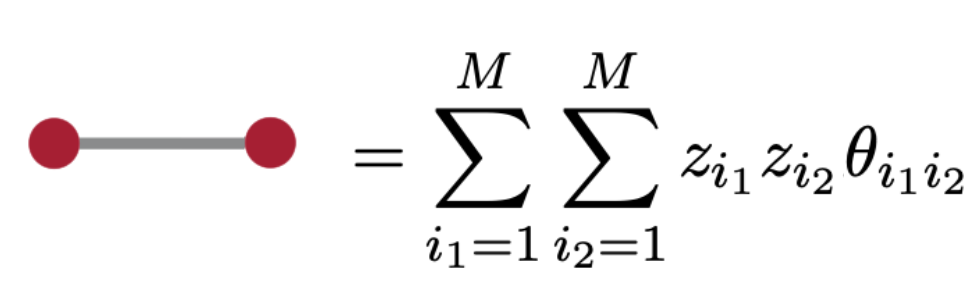}\\
  \caption{EFP construction: a degree one polynomial}
  \label{fig:deg1poly}
\end{figure}

Hence, two particles/constituents of a jet can be treated as two energy fractions with a single angularity connection between them, leading to a degree-1 polynomial, as can be seen from Figure~\ref{fig:deg1poly}

Because the EFP basis is infinite, a suitable organization and truncation scheme is necessary to use the basis in
practice. Several combinations of diagrams can be designed using the infinite number of vertex and particle correlator connections possible, however, the scope of this project has so far been restricted to exploring combinations with up to 7 particles/constituents/subjets and 8 angularity connectors between them. 

\begin{figure}[ht]
  \centering
  \includegraphics[width=0.8\textwidth]{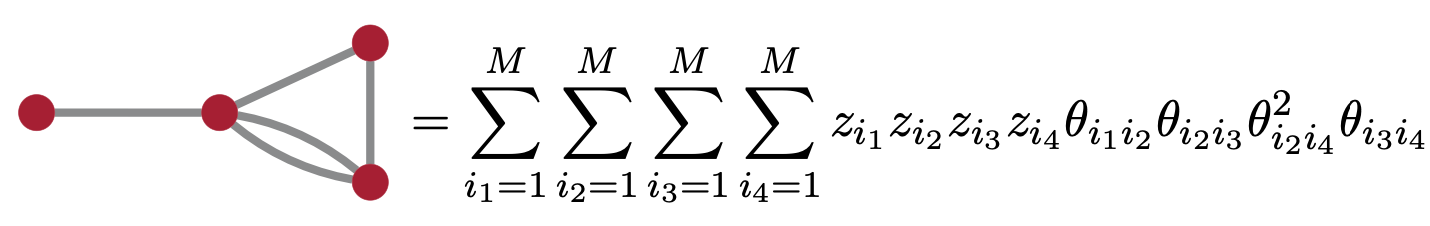}\\
  \caption{EFP diagram with 4 constituents and 5 angularity connectors}
  \label{fig:griddiag}
\end{figure}

The target is to implement EFPs in RIVET~\cite{Bierlich:2019rhm} framework and see if any particular combination of EFPs helps to distinguish between standard q/g jets and more unconventional jets. This might lead to a new jet-substructure observable for dark shower discrimination. Currently, a working setup (computing EFP multigraphs up to $N = 7$, $d = N - 1, N, N + 1$) is in place~\footnote{\url{https://gitlab.cern.ch/susinha/efp}} which takes into account the different possible orientations of the input “particles” and designs an array of possible EFP diagrams as a grid. For the EFP diagram shown in Figure~\ref{fig:griddiag}, a corresponding grid can be designed as follows in Table~\ref{tab:grid}, which translates the EFP to a set of "particle" pairs.

\begin{table}[tb]
\begin{center}
    \begin{tabular}{lcccc}
    \hline
       $N=4$, $d=N-2$ & 0 & 1  & 2 & 3\\
         \hline
        0  & - & - & - & - \\
        1  & 1 & - & 1 & 2 \\
        2  & - & - & - & - \\
        3  & - & - & 1 & - \\
      \hline
    \end{tabular}
    \end{center}
  \caption{Grid formation, translation Figure~\ref{fig:griddiag} into a set of "particle" pairs.}
  \label{tab:grid}
\end{table}

Individual EFP diagrams use $R=0.2$ anti-$k_T$~\cite{} subjets, from the leading jet in each event, as inputs and
the log-likelihood ratio (LLR) value between pseudodata (Signal + SM background) and MC (SM background) is 
obtained after a very inclusive particle-level analysis (no MET threshold, jet $p_\textrm{T} > \SI{50}{\GeV}$, 
$|\eta| <  4.9$, lepton vetoed).

\begin{figure}[ht]
  \centering
  \includegraphics[width=0.6\textwidth]{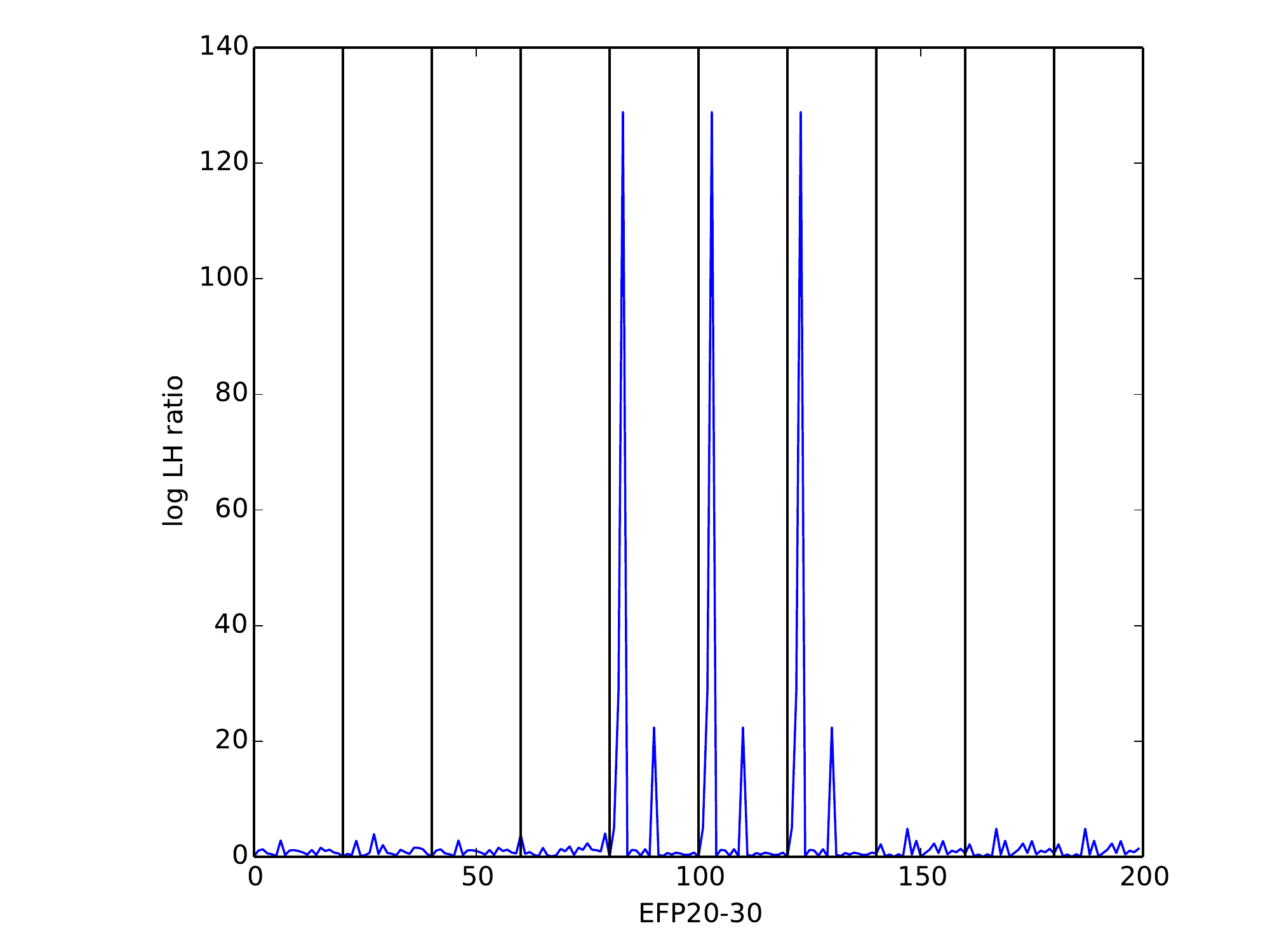}\\
  \caption{LLR summary distribution containing 10 EFP diagrams. showing a distinct Data vs MC difference in a few EFP diagrams (denoted by spikes).}
  \label{fig:llr}
\end{figure}

\begin{figure}[ht]
  \centering
  \includegraphics[width=0.35\textwidth]{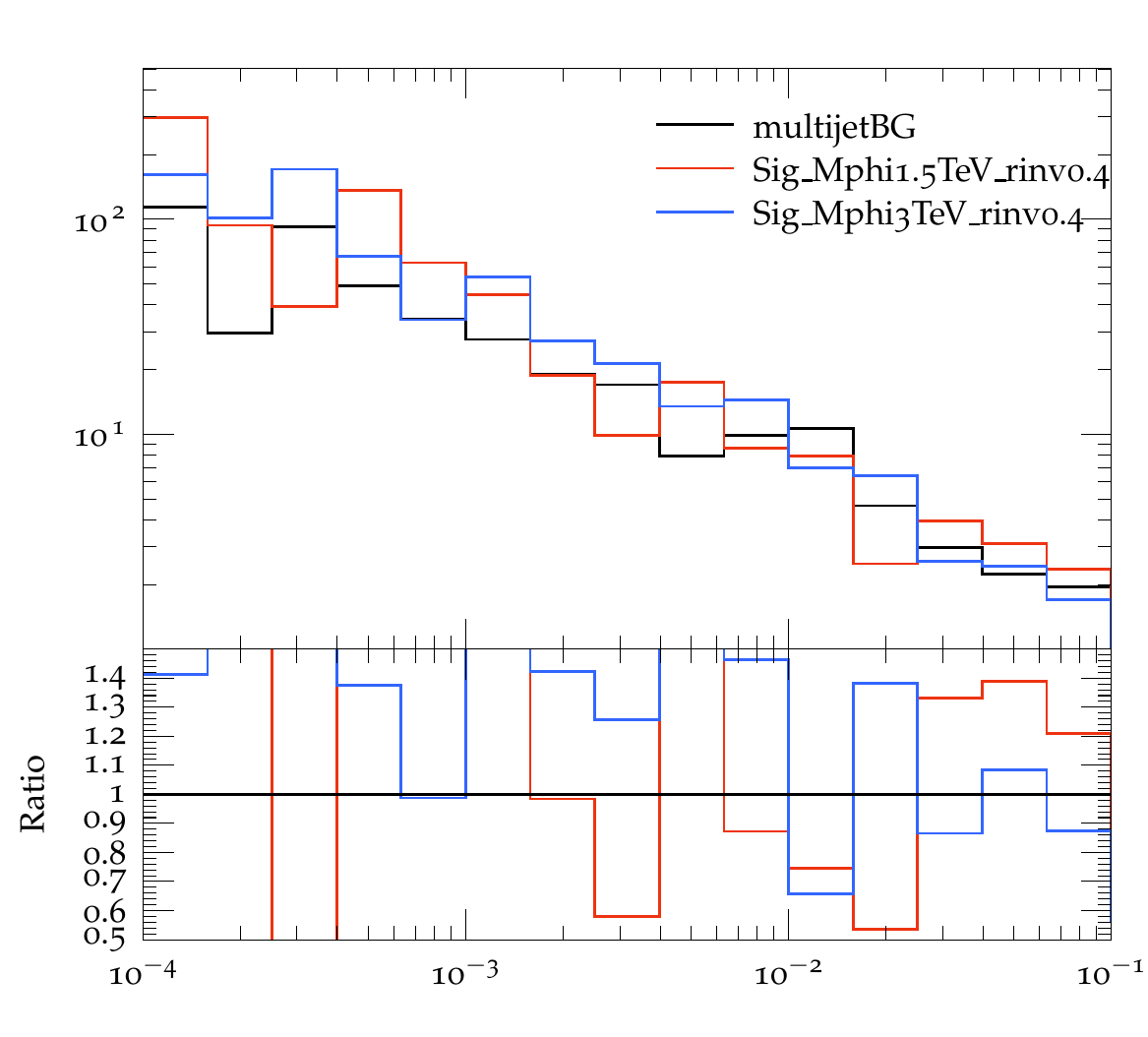}\qquad
  \includegraphics[width=0.35\textwidth]{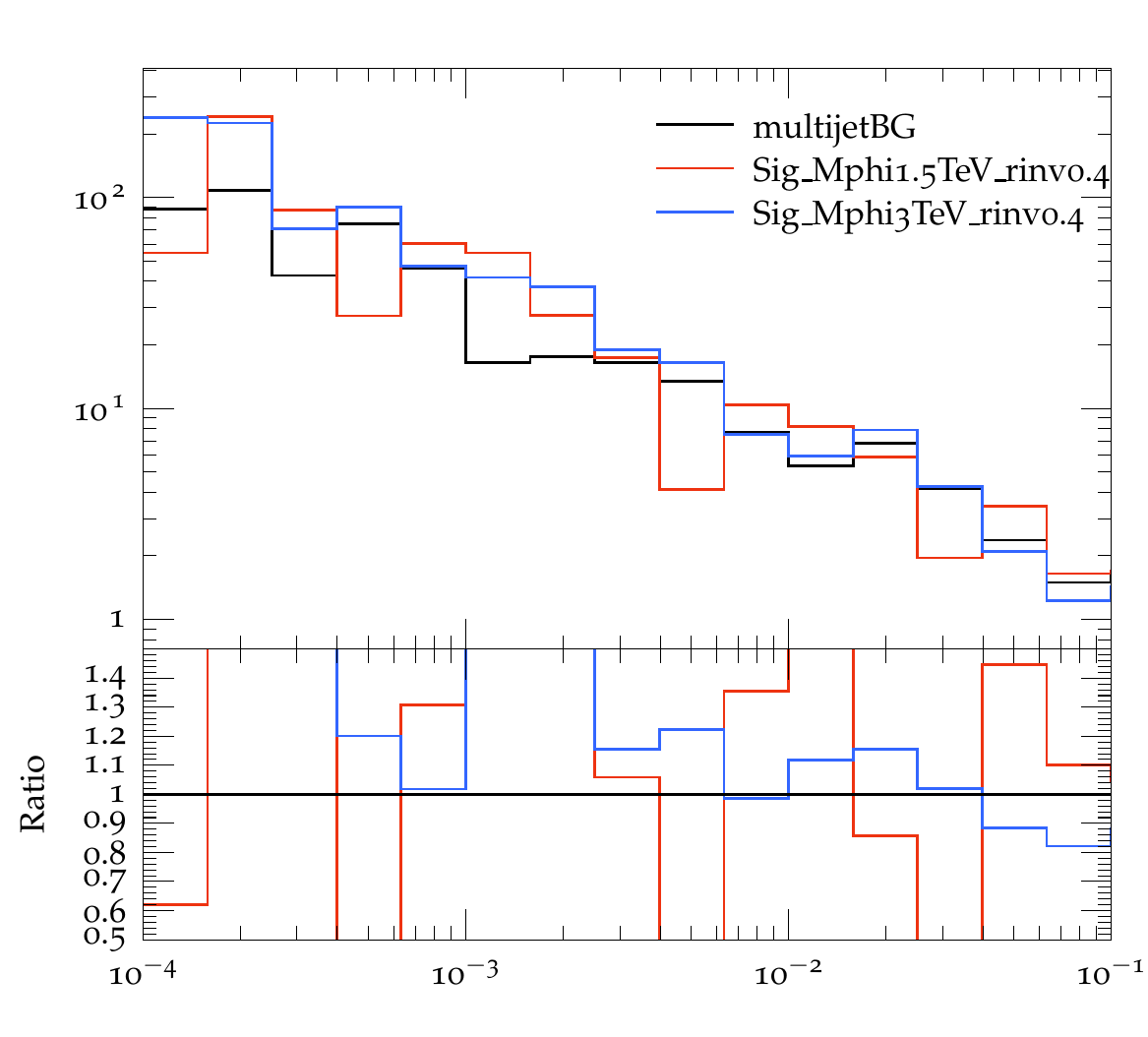}
  \caption{EFP distributions corresponding to spikes in LLR summary plot.}
  \label{fig:efpplot}
\end{figure}

Certain EFP diagrams seem to have some bins of the jet-shape observables that QCD just doesn't populate at all, in which the DM signal dominates. On comparing several of these LLR distributions, as can be seen from Figure~\ref{fig:llr}, some distinct LLRs have been identified that deviate from SM
(here, multijet BG is treated as a null-hypothesis), and the corresponding EFP diagrams are studied, as shown in Figure~\ref{fig:efpplot}. 

\section{Results}
\label{sec:res}
\begin{figure}[ht]
  \centering
  \includegraphics[width=0.45\textwidth]{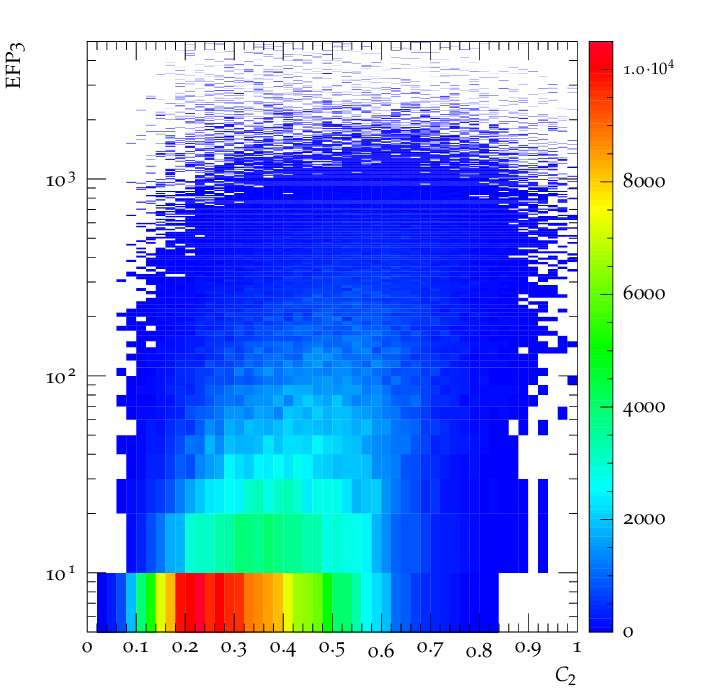}\qquad
  \includegraphics[width=0.45\textwidth]{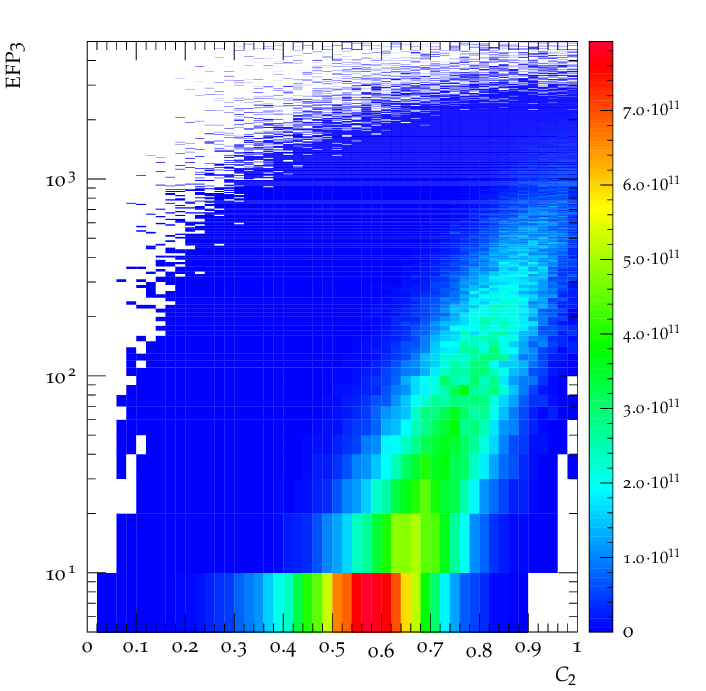}\\
  \includegraphics[width=0.45\textwidth]{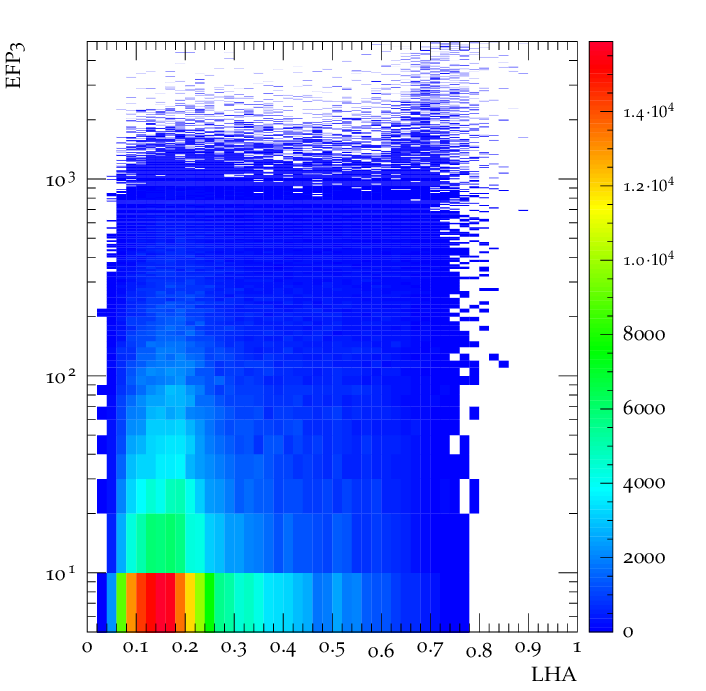}\qquad
  \includegraphics[width=0.45\textwidth]{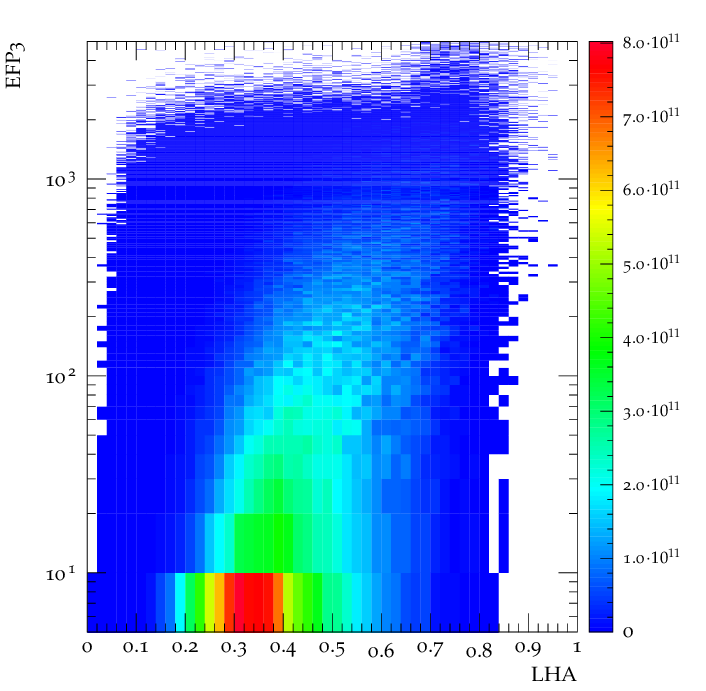}\\[1em]
  \caption{Comparisons of \CT\ (top), LHA (bottom) with respect to EFP3 between a signal corresponding to \rinv\ = 0.3, and \Mphi\ = 1.5 TeV (left) and the background (right).}
  \label{fig:jss2d}
\end{figure}

\begin{figure}[ht]
  \centering
  \includegraphics[width=0.45\textwidth]{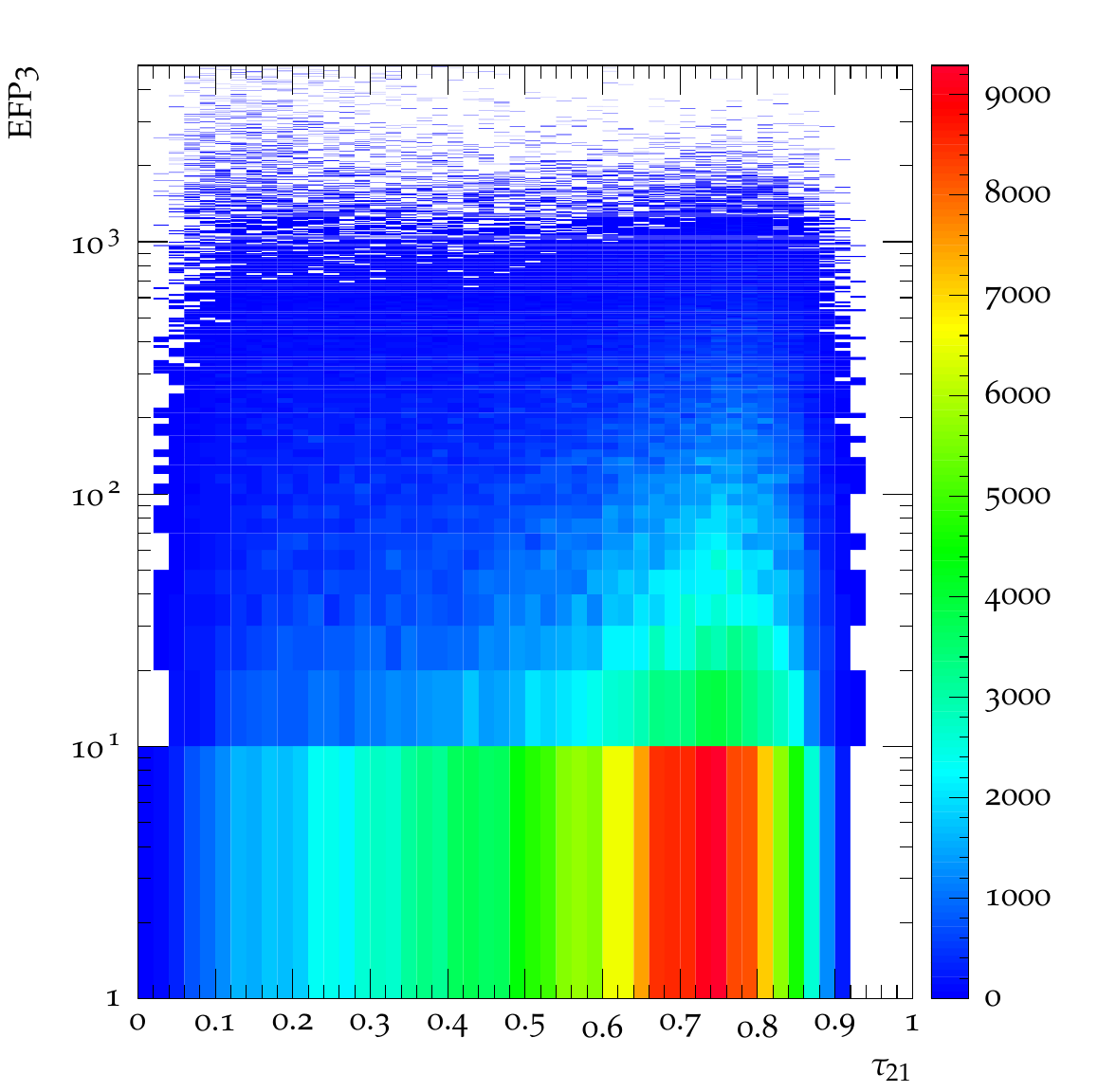}\qquad
  \includegraphics[width=0.45\textwidth]{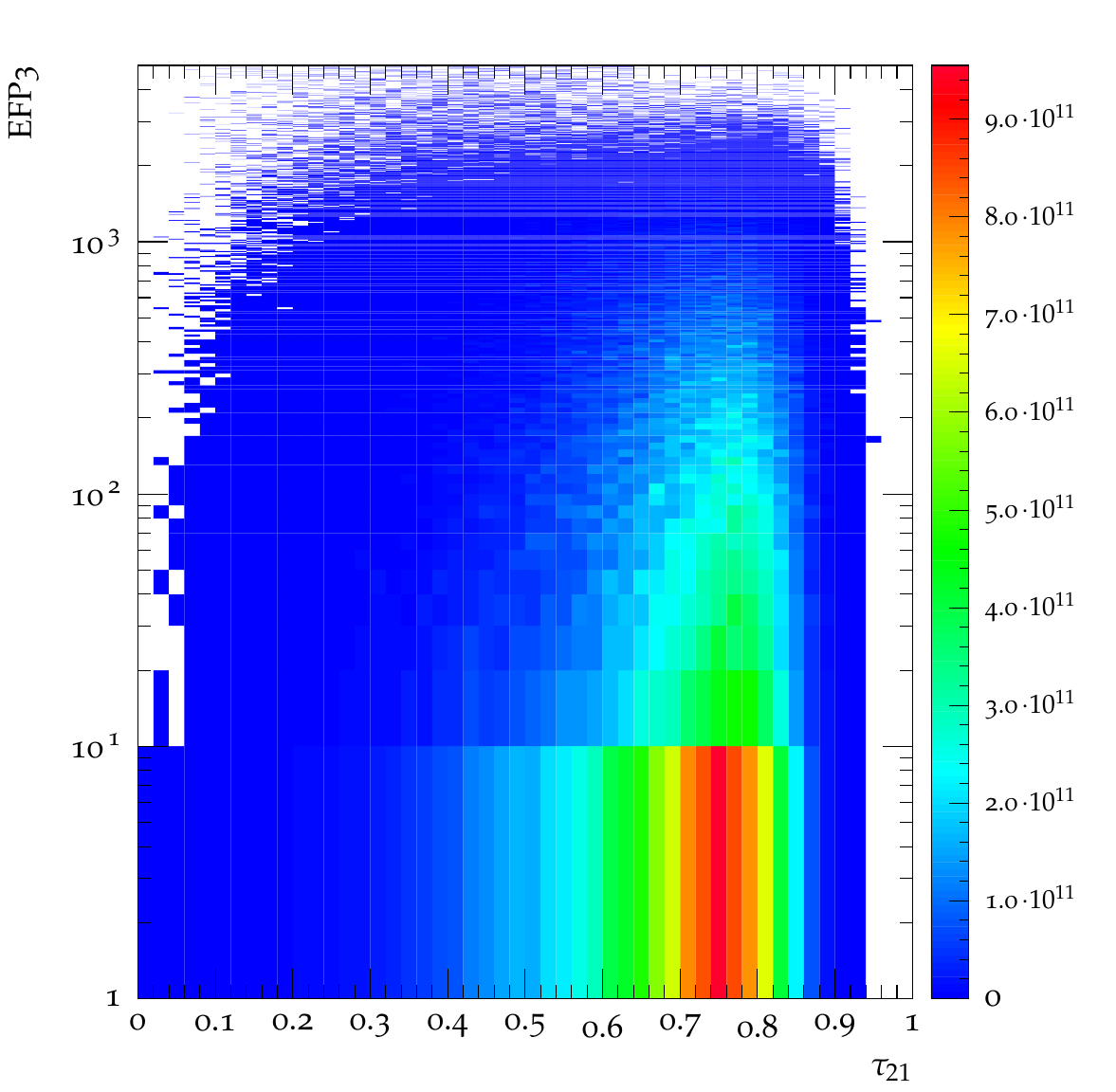}\\
  \includegraphics[width=0.45\textwidth]{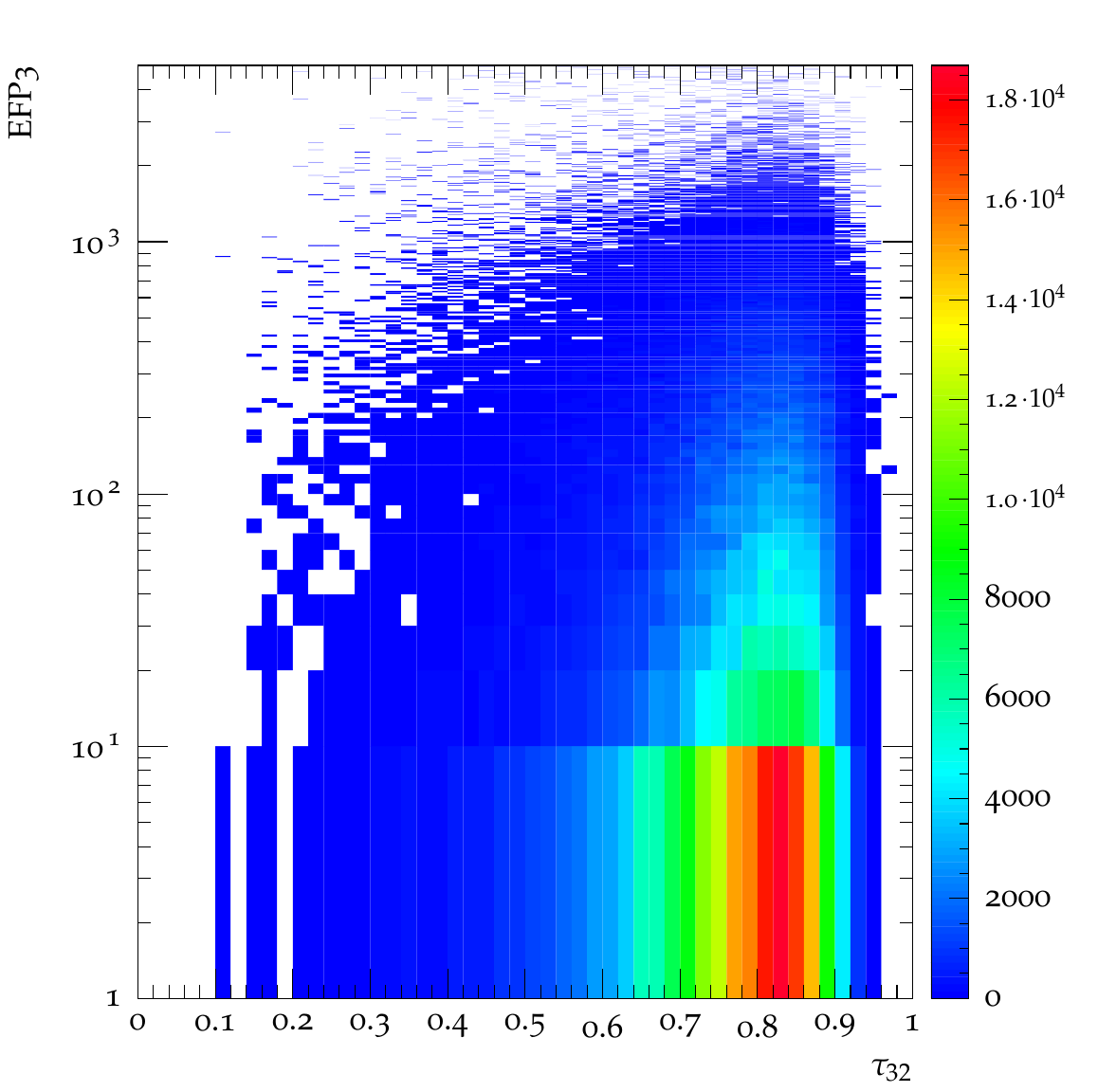}\qquad
  \includegraphics[width=0.45\textwidth]{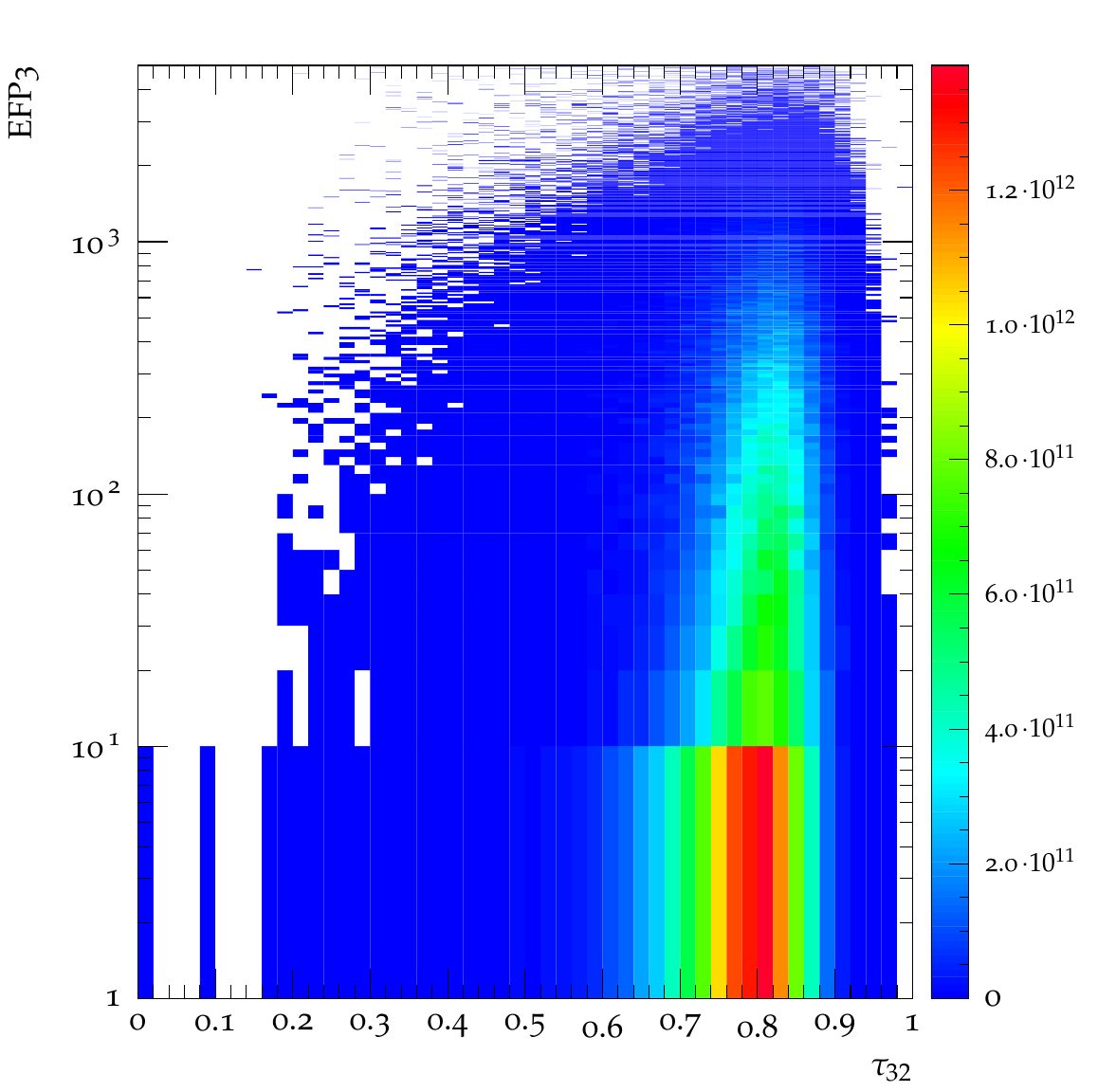}\\[1em]
  \caption{Comparisons of \tauTO\ (top), \tauTT\ (bottom) with respect to EFP3 between a signal corresponding to \rinv\ = 0.3, and \Mphi\ = 1.5 TeV (left) and the background (right).}
  \label{fig:jss2d1}
\end{figure}

It is necessary to understand what physics is being probed (i.e. the corresponding EFP equations)~\cite{Faucett:2022zie} and whether they are close to any standard jet substructure (JSS) variables. This is achieved by looking at correlations between the distinct EFPS and known JSS observables like n-subjettiness ratios~\cite{Thaler:2010tr}, energy correlation functions (single ratios~\cite{Larkoski:2013eya}), Les-Houches angularity (LHA)~\cite{Gras:2017jty}. 2D Distributions of several JSS observables vs a selected EFP are compared between semi-visible and ordinary jets in Fig.~\ref{fig:jss2d} and~\ref{fig:jss2d1}. In particular, it is observed that there are distinct populations of signal and background contributions at different ranges of \CT\ and LHA. The very preliminary results show that the correlation of EFPs with a standard JSS observable has the potential to provide improved discriminating power between the signal and background.

\section{Conclusion}
There are several avenues of strongly interacting dark sector that can be explored, specifically looking into unusual final state signatures. This preliminary study shows the possibility of probing these phase-space corners by exploiting the wealth of existing and new JSS observables in a IRC-safe linear basis, ala EFPs. Standalone EFPs or combination of several EFPs can be correlated to an existing JSS observable and lead to improved discriminating power between the signal and background. Next steps in the study involve providing concrete recommendations for the combinations that can be utlised in explorations of the dark-sector at the LHC.

\section*{Acknowledgements}
This work has received funding from the European Union's Horizon 2020 research and innovation programme as part of the Marie Skłodowska-Curie Innovative Training Network MCnetITN3 (grant agreement no. 722104).
AB also thanks STFC for support under Consolidated Grant ST/S000887/1. DK is funded by the National Research  Foundation (NRF), South Africa.
SS thanks University of Witwatersrand Research Council for Sellschop grant and NRF for Extension Support Doctoral Scholarship.

\bibliography{main.bib}

\nolinenumbers

\end{document}